\documentclass[preprint]{aastex}
\usepackage{emulateapj5}

\shorttitle{SZ Effect Scaling Relations}
\shortauthors{McCarthy et al.} 
\submitted{To appear in the Astrophysical Journal (received 10/02/02,
accepted 02/05/03)}

\begin{document} 

\title{Cluster Sunyaev-Zeldovich Effect Scaling Relations}

\author{Ian G. McCarthy$^{1,2}$, Arif Babul$^{1,3}$, Gilbert P. Holder$^4$, and Michael L. 
Balogh$^5$}

\affil{$^1$Department of Physics \& Astronomy, University of Victoria, Victoria, BC, 
V8P 1A1, Canada}

\affil{$^4$School of Natural Sciences, Institute for Advanced Study, Princeton, NJ, 
08540, USA}

\affil{$^5$Department of Physics, University of Durham, South Road, Durham, DH1 3LE, 
UK}

\footnotetext[2]{email: mccarthy@beluga.phys.uvic.ca}
\footnotetext[3]{CITA Senior Fellow}

\begin{abstract} 

X-ray observations of an ``entropy floor'' in nearby groups and clusters of galaxies 
offer evidence that important non-gravitational processes, such as radiative cooling 
and/or ``preheating'', have strongly influenced the evolution of the intracluster 
medium (ICM).  We examine how the presence of an entropy floor modifies the thermal 
Sunyaev-Zeldovich (SZ) effect.  A detailed analysis of scaling relations between X-ray 
and SZ effect observables and also between the two primary SZ effect observables is 
presented.  We find that relationships between the central Compton parameter and the 
temperature or mass of a cluster are extremely sensitive to the presence of an entropy 
floor.  The same is true for correlations between the integrated Compton parameter and 
the X-ray luminosity or the central Compton parameter.  In fact, if the entropy floor is 
as high as inferred in recent analyses of X-ray data, a comparison of these correlations 
with both {\it current} and future SZ effect observations should show a clear signature 
of this excess entropy.  Moreover, because the SZ effect is redshift-independent, the 
relations can potentially be used to track the evolution of the cluster gas and possibly 
discriminate between the possible sources of the excess entropy.  To facilitate 
comparisons with observations, we provide analytic fits to these scaling relations.

\end{abstract}


\section{INTRODUCTION}

Correlations between the global X-ray properties of galaxy clusters have proven to be 
important probes of the intracluster medium (ICM).  Studies of the relation between the 
X-ray luminosity ($L_X$) and the emission-weighted temperature ($T_X$) have been 
particularly powerful.  Numerical simulations and analytic models that take into account 
the effects of gravity and shock heating of the gas only (i.e., the so-called 
``self-similar'' models) predict that $L_X \propto T_X^2$ for massive clusters, yet the 
observed relation is much steeper; $L_X \propto T_X^{2.6-3.0}$ (e.g., Markevitch 
1998; Allen \& Fabian 1998; Arnaud \& Evrard 1999).  A number of other observed X-ray 
scaling relations, for example the total cluster mass ($M_{tot}$)-$T_X$ and total ICM 
mass  ($M_{gas}$)-$T_X$ relations, have also recently been shown to deviate from their 
predicted scalings (e.g., Horner et al. 1999; Ettori \& Fabian 1999; Mohr et al. 1999; 
Vikhlinin et al. 1999; Nevalainen et al. 2000; Finoguenov et al. 2001; McCarthy et al. 
2002, hereafter MBB02).  These discrepancies between theory and observations have 
motivated a number of authors to examine the potential role of ``additional'' gas 
physics.  For example, the heating of the ICM by galactic winds and/or quasar outflows 
has been investigated by Kaiser (1991) and also by a whole host of subsequent authors 
(e.g., Evrard \& Henry 1991; Bower 1997; Balogh et al. 1999; Wu et al. 2000; Loewenstein 
2000; Tozzi \& Norman 2001; Borgani et al. 2001; Babul et al. 2002, hereafter BBLP02; 
MBB02; Nath \& Roychowdhury 2002; Lloyd-Davies et al. 2002).  Recently, the effects of 
radiative cooling on X-ray scaling relations have also been explored (e.g., Bryan 2000; 
Voit \& Bryan 2001; Wu \& Xue 2002; Thomas et al. 2002; Voit et al. 2002; Dav\'{e} et al. 
2002).  These studies find that both heating and cooling can act in a similar manner, 
by raising the mean entropy of the intracluster gas and, in some cases, establishing a 
core in the entropy profile.  This, in turn, modifies the X-ray scaling relations of 
clusters and ameliorates, or possibly eliminates, the discrepancies between theory and 
observations.  It also potentially explains the emerging observational evidence for 
an ``entropy floor'' in nearby groups and low mass clusters (Ponman et al. 1999; 
Lloyd-Davies et al. 2000).

Thus far, X-ray observations alone have provided evidence for the entropy floor and 
it has come almost entirely from low redshift ($z \lesssim 0.2$) groups/clusters.  
Observations of higher redshift clusters are hindered by cosmological dimming [the 
bolometric X-ray surface brightness of a  cluster scales as $(1+z)^{-4}$].  An 
additional, {\it independent} probe which could be used to confirm the presence of this 
excess entropy in low/intermediate redshift clusters and also provide new tests for 
their high redshift counterparts would be quite useful.  Here, we show that scaling 
relations based on the thermal Sunyaev-Zeldovich effect (Sunyaev \& Zeldovich 1972; 
1980) - hereafter referred to as the ``SZ effect'' - can provide such a probe.

The SZ effect is a fractional change in the temperature/intensity of the cosmic microwave 
background (CMB) caused by the inverse-Compton scattering of CMB photons off high energy 
electrons in the ICM.  On average, the photons gain a small amount of this energy from 
the scatterings and this results in a slight spectral distortion of the CMB towards 
clusters.  At frequencies of $\nu \lesssim 218$ GHz, the SZ effect appears as a decrement 
in the temperature of the CMB, while at higher frequencies it appears as an increment.  
The magnitude of the SZ effect is determined by the integrated gas pressure along the 
line-of-sight through the cluster.  Since heating/cooling modifies the entropy of the 
gas, it should be expected that the gas pressure and, thus, the SZ effect will also be 
modified by these processes.  Therefore, it can be expected that SZ effect scaling 
relations will tell us something about the entropy history of the gas.  Unlike the X-ray 
surface brightness of a cluster, the SZ effect is not subject to cosmological dimming and 
can be used to study the effects of excess entropy in the ICM out to arbitrarily high 
redshift.  This makes scaling relations based on the SZ effect particularly interesting 
and attractive tests.

Traditionally, detecting and mapping the SZ effect has been quite difficult.  
This is because the effect is very weak, with typical beam-averaged decrements in the 
CMB temperature of only a few hundred $\mu$K (see Birkinshaw 1999 for a recent compilation).  
Recently, however, extraordinary leaps in detector technology and new observing strategies 
have allowed observers to make reliable and routine pointed observations of the effect (for 
recent data see, e.g., Grego et al.  2001; Grainge et al. 2002; Reese et al. 2002).  With 
this advance, we feel a detailed study of SZ effect scaling relations and what new insights 
they can give us on the thermal and spatial characteristics of the ICM, especially in high 
redshift clusters, is timely.

In this paper, we construct SZ effect scaling relations for massive clusters both with and 
without excess entropy and focus on how these correlations are modified by the presence of an 
entropy floor.  Along the way, we also discuss the potential for both current and future 
observations to measure these relations and constrain the entropy floor level.  
Constraining the entropy floor level out to high redshift could possibly yield 
information about the source(s) of the excess entropy, whether it be ``preheating'' by 
quasar outflows, radiative cooling, or some other non-gravitational process.  In a 
companion paper (McCarthy et al. 2003), we compare the scaling relations derived here to 
high redshift SZ effect data from the literature.

To date, only a few other theoretical studies have investigated the role of excess entropy 
on the SZ effect.  White et al. (2002), Springel et al. (2001), da Silva et al. (2001), 
Cavaliere \& Menci (2001), and Holder \& Carlstrom (2001) have all looked at how {\it 
universal} SZ effect properties, such as the SZ effect angular power 
spectrum, SZ effect cluster source counts, and/or the mean universal Compton parameter, 
are modified by an entropy floor.  However, measurements of most of these quantities 
are not feasible with current instrumentation and, therefore, estimates of the entropy 
floor level of clusters via observations of universal SZ effect quantities will have to 
wait until the next generation of instruments [e.g., the Sunyaev-Zeldovich Array ({\it 
SZA}), the Arcminute MicroKelvin Imager ({\it AMI}), and the Array for Microwave 
Background Anisotropy ({\it AMiBA})] come online.  In a spirit similar to that of the 
present study, however, Holder \& Carlstrom (2001) and Cavaliere \& Menci (2001) have 
also examined a few SZ effect scaling relations for {\it individual} clusters.  We 
briefly compare our findings to the results of these two studies in \S 5. 

The present paper is outlined as follows: in \S 2, we briefly describe the analytic 
cluster models developed in BBLP02 and employed here; in \S 3, we discuss in a general 
sense how and why entropy injection is expected to influence the SZ effect and how this 
can be explored with current and future observations through comparisons with predicted 
scaling relations.  In \S 4, we derive theoretical scaling relations based on SZ effect 
observables and quantify how excess entropy modifies these relations.  We start by 
examining scaling relations consisting only of SZ effect observables.  These relations 
are most interesting because they potentially offer a completely (X-ray-)independent 
way of probing the intracluster gas.  Next, we explore scaling relations between the 
various SZ effect observables and a cluster's mass.  These relations, too, can potentially 
be measured independent of X-ray observations, since clusters can be weighed via 
gravitational lensing or galaxy velocity dispersion measurements.  Finally, we construct 
and analyse scaling relations between the various SZ effect and X-ray observables (i.e., 
X-ray luminosity and emission-weighted temperature).  In \S 5, we discuss and summarize 
our findings.

The models we consider below were developed in a flat $\Lambda$-CDM cosmology with 
$h = 0.75$, $\Omega_m = 0.3$, and $\Omega_b = 0.020 h^{-2}$ (Burles et al. 2001).  They are 
computed for a number of different entropy floor levels spanning the range $K_0 \approx 100 
- 430$ keV cm$^2$.  This is approximately the range found to match the observed $L_X-T_X$ 
relations of both groups and hot clusters (e.g., Ponman et al. 1999; Lloyd-Davies et al. 
2000; Tozzi \& Norman 2001; BBLP02; MBB02).  We work in physical units 
throughout the paper (e.g., $M_{\odot}$ rather than $h^{-1} M_{\odot}$).    

\section{Galaxy Cluster Models}

Since an in-depth discussion of the cluster models can be found in BBLP02, we present 
only a brief description of the models here.  We note that the model clusters derived 
here represent high mass ($M_{tot} \gtrsim 3 \times 10^{14}$ M$_{\odot}$), high
temperature ($T_X \gtrsim 3$ keV) clusters only.  Such a range reflects all of the 
clusters observed to date through the SZ effect and is approximately the range expected 
to be probed by upcoming interferometric surveys (e.g., Holder et al. 2000).

\subsection{The Self-Similar Model}

To mimic the standard self-similar result deduced from numerical simulations, we 
implement the ``isothermal'' model of BBLP02.  In this model, the ICM is assumed to be 
isothermal and is in hydrostatic equilibrium with the dark halo potential.  The 
distribution of the dark matter in these clusters is assumed to be the same as found 
in recent high resolution numerical simulations (Moore et al. 1998; Lewis et al. 2000).  
This model will serve as our ``baseline'' model for assessing how an 
entropy floor modifies the SZ effect.  The isothermal model has been tested extensively 
(BBLP02; MBB02) to make sure it provides a good match to the results of 
genuine self-similar numerical simulations, such as those done by Evrard et al. (1996).  

\subsection{The Entropy Floor Models}

To test the effects of an entropy floor on the SZ effect, we make use of the entropy 
injection (preheated) models of BBLP02.  The models can be summarized as follows:  the 
dominant dark matter component, which is unaffected by the energy injection, collapses 
and virializes to form bound halos.  The distribution of the dark matter in such halos 
is assumed to be the same as for the self-similar clusters described above.  While the 
dark component is unaffected by energy injection, the collapse of the baryonic 
component is hindered by the pressure forces induced by entropy injection.  If the 
maximum infall velocity due purely to gravity of the dark halo is subsonic, the flow 
will be strongly affected by the pressure and it will not undergo accretion shocks.  It 
is assumed that the baryons will accumulate onto the halos {\it isentropically} at the 
adiabatic Bondi accretion rate (as described in Balogh et al. 1999).  This treatment, 
however, is only appropriate for low mass halos.  If the gravity of the dark halos is 
strong enough (as it is expected to be in the hot clusters being considered here) that 
the maximum infall velocity is transonic or supersonic, the gas will experience an 
additional (generally dominant) entropy increase due to accretion shocks.  In order to 
trace the shock history of the gas, a detailed knowledge of the merger history of the 
cluster/group is required but is not considered by BBLP02.  Instead, it is assumed that at 
some earlier time the most massive cluster progenitor will have had a mass low enough such 
that shocks were negligible in its formation, similar to the low mass halos discussed 
above.  This progenitor forms an isentropic gas core of radius $r_{c}$ at the cluster 
center.  The entropy of gas outside of the core, however, will be affected by shocks.  
Recent high resolution numerical simulations suggest that the ``entropy'' profile for gas 
outside this core can be adequately represented by a simple analytic expression given by 
$\ln{K(r)} = \ln{K_0} + \alpha \ln{(r/r_c)}$ (Lewis et al. 2000), where $K \equiv 
kT_e n_e^{-2/3}$.  For the massive, hot clusters ($T_X \gtrsim 3$ keV)
of interest here, $\alpha \sim 1.1$ (Tozzi \& Norman 2001; BBLP02).
 
Following this prescription and specifying the parameters $r_c$, $\rho_{gas}(r_c)$,
and $\alpha$ (as discussed in BBLP02) completely determines the models.  Under all
conditions, the gas is assumed to be in hydrostatic equilibrium within the dark halo 
potential.  The complicated effects of radiative cooling are neglected by these models. 

\section{Entropy Injection and the SZ Effect}

The amplitude of the SZ effect is directly proportional to the ``Compton parameter'' 
($y$) which is given by

\begin{equation}
y(\theta) = \frac{\sigma_{T}}{m_e c^2} \int P_e(\vec{r}) dl
\end{equation}
 
\noindent where $\theta$ is the projected position from the cluster center, $\sigma_T$ is 
the Thomson cross-section, and $P_e(\vec{r}) \equiv n_e(\vec{r}) kT_e(\vec{r})$ is the 
electron pressure of the ICM at the 3-dimensional position $\vec{r}$.  The integral 
is performed over the line-of-sight ($l$) through the cluster.

All of the physics of the SZ effect is contained within the Compton parameter.  It 
is the SZ effect analog of the X-ray surface brightness of a cluster and is a measure 
of the average fractional energy gain of a photon due to inverse-Compton scattering 
while passing through a cloud of gas (in this case, the ICM) with an electron pressure 
profile of $P_e(\vec{r})$.  As discussed by BBLP02 and MBB02, the presence of excess 
entropy will modify both a cluster's density and temperature profiles.  In the case 
where it is preheating 
{\epsscale{1.0}
\plotone{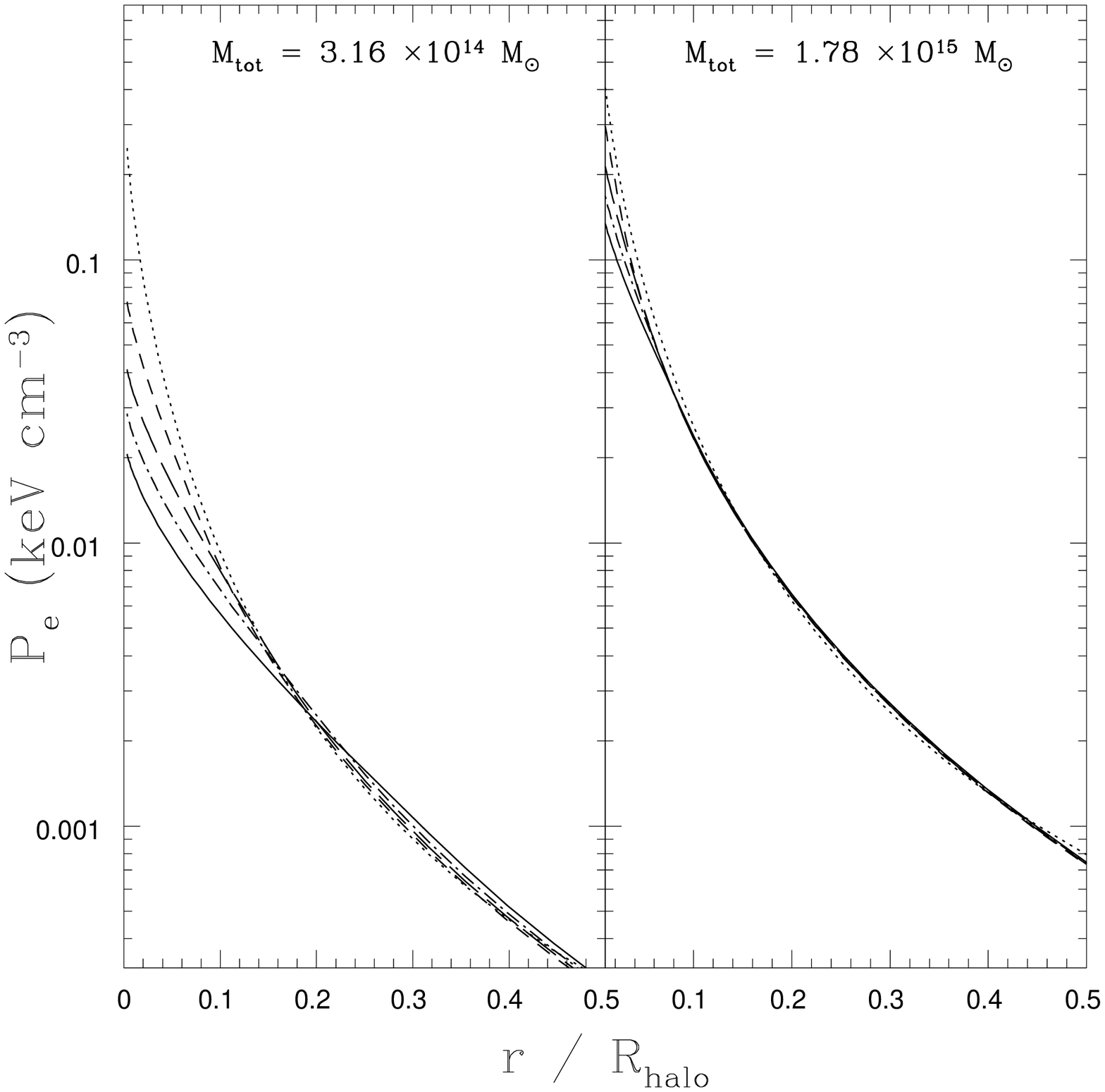}
{Fig. 1. \footnotesize
Effects of an entropy floor on cluster pressure profiles.  {\it Left:} Cluster
with total mass $M_{tot} = 3.16 \times 10^{14} M_{\odot}$.  {\it Right:} Cluster with
total mass $M_{tot} = 1.78 \times 10^{15} M_{\odot}$.  The dotted line is the standard
self-similar result.  The short-dashed, long-dashed, dot-dashed, and solid
lines represent the models of BBLP02 with entropy floor constants of $K_0$ = 100, 200,
300, and 427 keV cm$^2$, respectively.
}}
\vskip0.1in
\noindent
that gives rise to an entropy core, as in the present study, the 
temperature of the gas near the center of the cluster is increased and, therefore, so is 
the global emission-weighted temperature of the cluster (e.g., Fig. 1 of MBB02).  At the 
same time, the density of the gas at the cluster center is dramatically reduced (e.g., 
Fig. 2 of MBB02).  It turns out that, relatively speaking, preheating has a stronger 
influence on the density than it does on the temperature, at least at the centers of 
massive clusters.  The result is that the gas pressure in central regions of a cluster is 
reduced by preheating and, consequently, so is the cluster's Compton parameter. To 
demonstrate this, we plot cluster pressure profiles ($z = 0.2$) for several values of the 
entropy floor in Figure 1 ($R_{halo}$ is the radius of the cluster).  The 
addition of an entropy floor leads to a decrease in the gas pressure near the cluster core.  
The gas pressure in the outer regions of the clusters, however, remains relatively unchanged 
as the entropy increase due to gravitational shock heating dominates the non-gravitational 
entropy injection.  Also of note is that the {\it relative} difference between the various 
models is greatest for the lower mass cluster.  This is expected since the lower mass cluster 
has a  shallower potential well and, thus, is more strongly influenced by the presence of an 
entropy floor.

With an entropy floor significantly affecting the pressure of the ICM near the center 
of a cluster, the Compton parameter will be most strongly modified if it is evaluated 
within the smallest possible projected radius [i.e., the {\it central} Compton 
parameter, $y(\theta = 0) \equiv y_0$].  Integrating (or averaging) the Compton 
parameter within larger projected radii (for example, $R_{halo}$, the radius of the 
cluster), on the other hand, will diminish (though not completely remove) the effects 
of entropy injection.  Which of these quantities, the central or integrated Compton 
parameters, will be the most useful for constructing scaling relations that are 
sensitive to the value of $K_0$ will depend upon which other cluster observables are 
used in the relations.  For example, in \S 4, we show that a comparison of $y_0$ with 
$T_X$ provides a more sensitive test of entropy injection than a comparison of the 
integrated Compton parameter with $T_X$.  On the other hand, a scaling relation between 
the integrated Compton parameter and $L_X$ is more sensitive to the entropy floor level 
than is a scaling relation between $y_0$ and $L_X$.  Ultimately, one would like to directly 
image the entire cluster out to the virial radius with high angular resolution and directly 
compare theoretical models to the images over the entire cluster profile.

The observability of these quantities will obviously depend on the details of the 
instrument and observing strategy.  Generically, 
observations of the SZ effect filter large-scale emission while finite resolution 
smears out small-scale structures.  Fitting to a model (such as the isothermal $\beta$ 
model; Cavaliere \& Fusco-Femiano 1976; 1978) provides a method for effectively 
deconvolving these effects and estimating the central and integrated Compton 
parameters, but it is important to keep in mind that such quantities are inferred and 
model-dependent.  Provided the smallest angular scale resolved is comparable to the 
typical scale over which the cluster varies, the inferred $y_0$ will be reliable, 
while inferred integrated Compton parameters will not be reliable when extrapolated 
beyond the filtering scale of the observations.  For current interferometric 
observations [such as those obtained with the Berkeley Illinois Maryland Association 
({\it BIMA}) and Owens Valley Radio Observatory ({\it OVRO}) arrays and the Ryle 
telescope], the highest angular resolution for SZ measurements is typically smaller 
than the core radius of most clusters observed to date ($\sim$ 30'') while the typical 
large-scale filtering becomes important on scales larger than about 2'. Therefore, it can 
be expected that the inferred values of $y_0$ should be reasonably accurate while any 
integrated Compton parameter/flux density reported on scales larger than 2' will be 
suspect.

In recognition of the above, we construct scaling relations between central Compton 
parameter, the integrated Compton parameter within the central 1' (which is a 
conservative choice for the filtering scale of current interferometers), and various 
X-ray observables.  However, we also explore relations involving the integrated 
Compton parameter within $R_{halo}$, as future experiments, such as {\it SZA}, 
will probe angular scales larger than that of current experiments.

\section{SZ Effect Scaling Relations}

\subsection{100\% SZ Effect: The $S_{\nu}-y_0$ Relation}

We start our discussion of SZ effect scaling relations by first examining relations 
between SZ effect quantities only (i.e., the central and integrated Compton 
parameters).  In principle, these relations are completely independent of X-ray 
results, in that the latter depend on the combination of $n_e^2 \epsilon(T_e)$ [where 
$\epsilon(T_e)$ is the X-ray emissivity] whereas the SZ effect depends on $n_e T_e$.  
For this reason, scaling relations which depend only on SZ effect quantities have the 
potential of providing new insights into both the thermal and spatial properties of the 
ICM. As such, we will invest a little bit more effort discussing and expanding on the 
nature of these relations than the SZ effect-X-ray relations discussed later in \S4.2-3.

First, because the SZ effect flux density of a cluster is most commonly reported in 
observational studies and not the integrated Compton parameter, we convert the 
integrated Compton parameter (symbolized as $y_{int}$), which is given by

\begin{equation}
y_{int}(\leq \theta) = 2 \pi \int_0^{\theta} y(\theta') \theta' d\theta'
\end{equation}
 
\noindent into flux density ($S_{\nu}$) via

\begin{equation}
S_{\nu} = j_{\nu}(x) y_{int}
\end{equation}

\noindent where $j_{\nu}(x)$ describes the shape of the SZ effect spectrum and is a 
function of the dimensionless frequency $x = h \nu / k T_{CMB}$.  The mean temperature 
of the present-day cosmic microwave background, $T_{CMB}$, is 2.728 K (Fixsen et al. 
1996) and $j_{\nu}(x) = 2(kT_{CMB})^3 (hc)^{-2} f_{\nu}(x)$, with

\begin{equation}
f_{\nu}(x) = \frac{x^4 e^x}{(e^x - 1)^2} \biggr( \frac{x}{\tanh(x/2)} -4 \biggl)
\end{equation}

\noindent where $f_{\nu} \approx -2x^2$ at long wavelengths (the Rayleigh-Jeans limit).

Since our aim here is to achieve a {\it physical} understanding of the relationship 
between $S_{\nu}$ and $y_0$, we remove the frequency dependence of the integrated SZ 
effect flux density by dividing it by the quantity $f_{\nu}$ given in (4).  Then a 
comparison of the quantity $S_{\nu}/f_{\nu}$ with genuine observations at any observing 
frequency $\nu$ can be made simply by multiplying by the conversion factor $f_{\nu}$.   
For reasons discussed in \S 3, we investigate the frequency-independent SZ effect flux 
density within a fixed angular radius of $1\arcmin$ (which we symbolize as 
$S_{\nu,arc}/f_{\nu}$) and within $R_{halo}$, the radius of the cluster (which we 
symbolize as $S_{\nu,halo}/f_{\nu}$).  

In Figure 2, we plot scaling relations between the (frequency-independent) SZ effect 
flux densities and the central Compton parameter.  The dotted lines are the 
self-similar results.  The short-dashed, long-dashed, dot-dashed, and solid lines 
represent the results of the models of BBLP02 with entropy floor constants of $K_0$ = 
100, 200, 300, and 430 keV cm$^2$, respectively.  The top panel is for the SZ effect 
flux density within central one arcminute, while the bottom panel is the total SZ effect 
flux density of the cluster (out to $R_{halo}$).  The thick lines are for $z = 0.2$ and 
the thin lines are for $z = 1.0$.

We start by considering the dependence of $S_{\nu,arc}/f_{\nu}-y_0$ on $K_0$ and $z$ 
in the top panel of Figure 2.  First, as discussed in \S 3, entropy injection leads 
to a decrease in the gas pressure near the cluster core (Fig. 1).  This, in turn, reduces 
the magnitude of both $S_{\nu,arc}/f_{\nu}$ and $y_0$.  However, the relative reduction 
in both of these quantities is not equal.  $S_{\nu,arc}/f_{\nu}$ is derived within 
larger physical radii than $y_0$ and, therefore, is less affected by the entropy floor.  
The result is an increase in the normalization of the $S_{\nu,arc}/f_{\nu}-y_0$ 
relation as the value of $K_0$ is increased.  For example, at $z = 0.2$, a cluster with 
a fixed value $\log{y_0} \approx -3.7$ ($\Delta T_0 \approx$ -1060 $\mu K$ at 30 GHz) 
will have an integrated decrement of  $S_{\nu,arc}/f_{\nu} \approx 5.5$ mJy ($S_{\nu} 
\approx -11$ mJy at 30 GHz) if no entropy floor is present but will 
{\epsscale{1.0}
\plotone{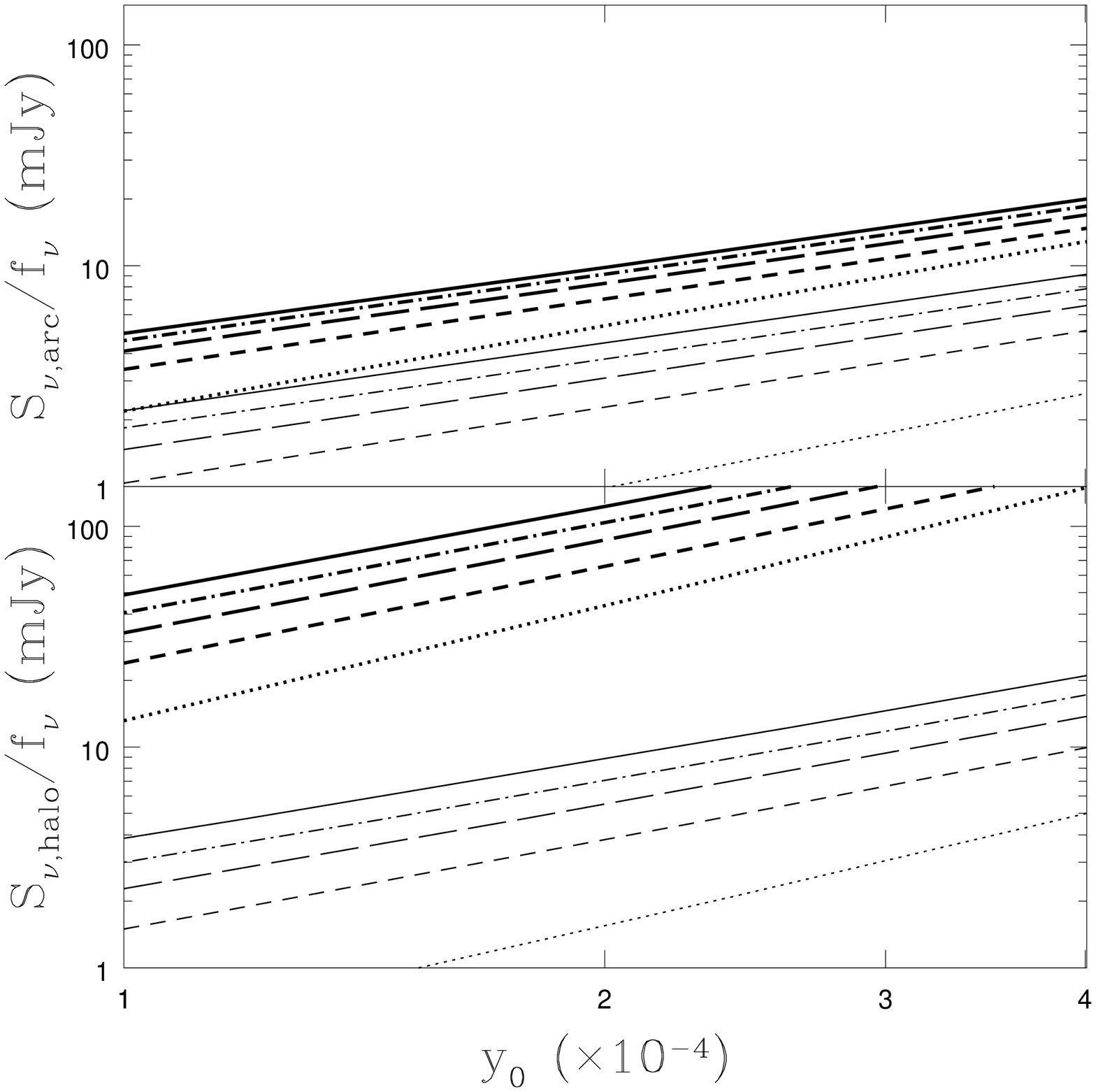}
{Fig. 2. \footnotesize
Comparison of the $S_{\nu} / f_{\nu} -  y_0$ relations.  The short-dashed,
long-dashed, dot-dashed, and solid lines represent the models of BBLP02 with entropy floor
levels $K_0$ = 100, 200, 300, and 427 keV cm$^2$, respectively.  The dotted lines are the
isothermal self-similar result.  The thick lines are for $z = 0.2$ and the thin lines
are for $z = 1.0$.  The top panel is for the SZ effect flux density within the central one
arcminute and the bottom panel is for the total SZ effect flux density of the
cluster.
}}
\vskip0.1in
\noindent
have $S_{\nu,arc}/f_{\nu} \approx 10$ mJy ($S_{\nu} \approx -20$ mJy at 30 GHz) if $K_0 
\approx 430$ keV cm$^2$.  Furthermore, this trend is amplified at higher redshifts 
since the SZ effect flux density within a fixed angular size probes larger physical 
regions at higher redshifts (and, therefore, is less affected by the entropy floor).  
Compare the difference between the models for the $z = 0.2$ lines and the $z = 1.0$, 
for example.  

The slope of $S_{\nu,arc}/f_{\nu}-y_0$ relation is also modified by the presence of an 
entropy floor.  For example$^6$, at $z = 0.2$ and 1.0 (respectively), the self-similar 
model approximately predicts 

\footnotetext[6]{Throughout the paper we often compare scaling relations at $z$ = 0.2 
and 1.0 for the self-similar and $K_0 = 300$ keV cm$^2$ models.  The choice in redshift 
is motivated by the fact that most interferometric observations are of clusters with $0.2 
\lesssim z \lesssim 1.0$, while the choice in entropy floor level is motivated by X-ray 
observations which require $K_0 \gtrsim 300$ keV cm$^2$ for massive clusters.}

\begin{equation}
S_{\nu,arc}/f_{\nu} \propto y_0^{1.3,~1.4}
\end{equation}

\noindent while the $K_0 = 300$ keV cm$^2$ model approximately predicts 

\begin{equation}
S_{\nu,arc}/f_{\nu} \propto y_0^{1.0,~1.1}
\end{equation}

\noindent A more precise analytic expression for the $S_{\nu,arc}/f_{\nu}-y_0$ 
relation at any $z \lesssim 1$ and any entropy floor in the range $K_0 \approx 100 - 700$ 
keV cm$^2$ is presented in Table 1.  

At a fixed redshift, the presence of an entropy floor flattens the relationship between 
$S_{\nu,arc}/f_{\nu}$ and $y_0$ because clusters with small values of $y_0$ (i.e., low 
mass clusters) are more strongly influenced (relatively speaking) by entropy injection 
than clusters with large values of $y_0$.  

It is worth noting that, for the most part, the $S_{\nu,arc}-y_0$ relations (and, in 
fact, virtually every relation discussed in the paper) exhibit almost perfect power-law 
behavior.  This may seem rather surprising in light of the fact that most X-ray scaling 
relations show some kind of a break from power-law trends (see BBLP02, for example).  We 
note, however, that the break in the $L_X-T_X$ relation, for example, occurs at the 
transition between groups and clusters (roughly $T_X \lesssim 1$ keV).  This is below the 
range of temperatures (and central Compton parameters) studied in the present paper.  We 
verify that a break does occur for low mass systems (i.e., for systems with $y_0 \lesssim 
10^{-5}$) but, because such low mass systems have very weak integrated SZ effect signals, 
there is little hope of observing this feature in the near future.

In the bottom panel of Figure 2, we plot the $S_{\nu,halo}/f_{\nu}-y_0$ relations.  
With the quantity $S_{\nu,halo}/f_{\nu}$ only slightly affected by the presence of an 
entropy floor (and, therefore, the relative difference in the normalizations of the 
models are at a maximum), these relations are more sensitive to entropy injection 
than those plotted in the top panel of Figure 2.  For example, for a $z = 0.2$ cluster 
with a fixed value of $\log{y_0} \approx -3.7$, the $K_0 = 430$ keV cm$^2$ models 
predicts a flux density that is roughly three times larger than that predicted by the 
self-similar model (as opposed to being only about two times larger for the relation 
presented in the top panel).  

The steepness of the relation is also altered by an entropy floor.  Using simple 
scaling arguments, one can easily arrive at the predicted self-similar result.  
Combining $S_{\nu,halo}/f_{\nu} \propto T_X R^3$ and $y_0 \propto T_X R$ with 
$R \propto T_X^{0.5}$ (which is just the virial theorem) yields 

\begin{equation}
S_{\nu,halo}/f_{\nu} \propto y_0^{5/3},
\end{equation}

\noindent in excellent agreement with the predictions of the BBLP02 
isothermal model.  However, injection of entropy flattens the relation and at $z = 0.2$ 
and 1.0, the $K_0 = 300$ keV cm$^2$ model approximately predicts

\begin{equation}
S_{\nu,halo}/f_{\nu} \propto y_0^{1.3,~ 1.2}
\end{equation}

\noindent An analytic expression for the $S_{\nu,halo}/f_{\nu}-y_0$ relation at any $z 
\lesssim 1$ and any entropy floor between 100 keV cm$^2$ $\lesssim K_0 \lesssim 700$ keV 
cm$^2$ is given in Table 1.

Since the $S_{\nu,halo}/f_{\nu}-y_0$ relation is more sensitive to entropy
injection (relatively speaking) than the $S_{\nu,arc}/f_{\nu}-y_0$ relation, this means that
observations obtained with future experiments, such as the {\it SZA}, {\it AMiBA}, and {\it
AMI} (which will probe larger angular scales than that of current interferometers), will
(at least in theory) have the best chance of placing tight constraints on the 
non-gravitational entropy of high redshift clusters.

We note that most of the dependence of the relations plotted in Figure 2 (both the 
top and bottom panels) on the entropy floor level, $K_0$, is due to the central Compton 
parameter, $y_0$.  A completely resolved central decrement formally requires infinite 
resolution and there is little to be gained by increasing the resolution beyond the smallest 
scale on which the cluster structure varies.  For diffuse emission, higher resolution 
normally means less signal per resolution element, so pushing to very high resolution will 
not lead to much new information.  Alternatively, the central Compton parameter can be 
estimated by fitting a model (e.g., the isothermal $\beta$ model) to the the observed SZ 
effect surface brightness and extrapolating it to the cluster center (as discussed briefly 
in \S 3).  This method is widely used to estimate $y_0$ (e.g., Carlstrom et al. 1996; 
Holzapfel et al. 1997; Grego et al. 2000; 2001; Reese et al. 2000; 2002; Pointecouteau et 
al. 2001; 2002; Jones et al. 2002; Grainge et al. 2002).  While the {\it statistical} 
measurement error on $y_0$ for current data is typically only of order $100$ $\mu$K (which 
is relatively small compared to the differences between the models plotted in Figure 2, 
assuming $S_{\nu}/f_{\nu}$ is known), it has yet to be demonstrated that the {\it 
systematic} error (due to assuming an incorrect surface brightness model) for current data 
is negligible.  If the models employed in the extrapolation provide poor descriptions of the 
surface brightness profiles, then one would expect the results to vary as a function of 
instrument characteristics (e.g., resolution, field of view).  However, a comparison of the 
results of various studies (which made use of different instruments; e.g., {\it Ryle} 
telescope, {\it BIMA/OVRO}, and {\it SuZIE}) of the same clusters reveals that the agreement 
is quite good, often within one sigma statistical uncertainty (see Holzapfel et al. 1997, 
Reese et al. 2002, and Jones et al. 2002, for example).  Thus, the extrapolation procedure 
seems to be a viable way estimating the central Compton parameter at current sensitivity.  By 
modeling ``mock'' (future) {\it SZA} observations, we also demonstrated that this 
extrapolation procedure is an accurate way of estimating the true underlying values of $y_0$ 
and $S_{\nu}$ of the BBLP02 cluster models (see McCarthy et al. 2003).

Finally, we point out that part of the motivation for our study of the $S_{\nu}-y_0$ 
relations (and the other scaling relations involving these quantities) comes from the 
fact that observational studies often use either $S_{\nu}$ or $y_0$ to characterize a 
cluster's SZ effect but not usually both.  For example, integrated SZ effect flux 
densities are often associated with large-beam single-dish experiments, while 
estimates of the central Compton parameter normally come from high resolution 
interferometric observations.  However, if one is able to determine $y(\theta)$ (i.e., the 
SZ effect ``surface brightness'' profile) from a single dataset then, strictly speaking, 
it isn't necessary to use scaling relations to probe the entropy of the ICM.  For 
example, the SZ effect surface brightness profile could be used together with the X-ray 
surface brightness profile (if it is known) to determine the true 3-dimensional entropy 
distribution of the gas (with some assumptions about the geometry of the cluster).  The 
disadvantages of this method are: (1) it requires X-ray observations, and (2) it probably 
cannot be used for the most distant clusters, since the X-ray signal-to-noise ratio falls 
sharply with increasing $z$.  Alternatively, the entropy distribution could be constrained 
by comparing the observed SZ effect surface brightness profile with theoretically 
predicted profiles.  This is similar to using scaling relations, since the entropy 
distribution is being inferred and not measured, but with the important difference that 
all of the available SZ effect information, $y(\theta)$, is being used in the comparison.  
We are in the process of exploring these methods and they will be addressed in detail in 
a subsequent paper.  For now, we stick with the scaling relations derived above, which 
are more readily comparable with available observations.

{\epsscale{1.0}
\plotone{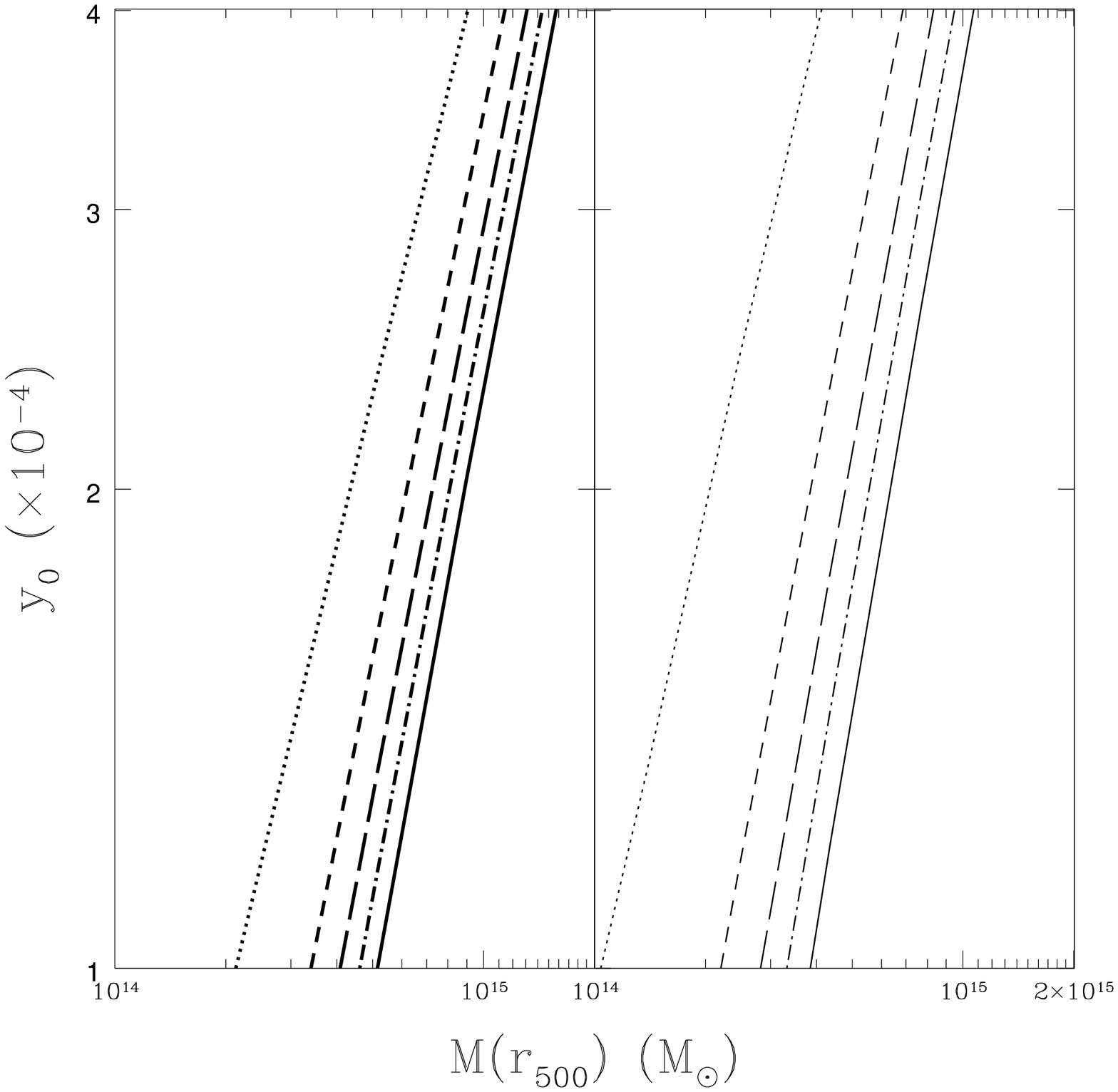}
{Fig. 3. \footnotesize
Comparison of the $y_0 - M(r_{500})$ relations.  The lines hold the same
meanings as in Fig. 2.  The left-hand panel is for $z = 0.2$ and the right-hand panel is
for $z = 1.0$.
}}
\vskip0.1in

\subsection{The $y_0$-X-ray scaling relations}

\subsubsection{$y_0-M(r_{500})$}

In Figure 3, we present scaling relations between $y_0$ and the total cluster dark 
matter mass within the radius $r_{500}$ [i.e., $M(r_{500})$].  This is the radius 
within which the mean dark matter mass density is 500 times the critical density at $z$ 
= 0.  The lines hold the same meaning as in Figure 2.  For clarity, we plot the $z = 
0.2$ predictions in the left-hand panel and the $z = 1.0$ predicts in the right-hand 
panel.

It is apparent that entropy injection has a substantial effect on the $y_0-M(r_{500})$ 
relation.  Both the normalization and the steepness of the relation are modified.  
First, the normalizations imply that, for a cluster of given mass at a given redshift, 
entropy injection tends to diminish the strength of $y_0$.  For example, at $z = 0.2$, a 
cluster with $M(r_{500}) \approx 9 \times 10^{14} M_{\odot}$ will have $y_0 \approx 4 
\times 10^{-4}$ in the absence of an entropy floor, but the central Compton parameter is 
only half this value for an entropy floor of $K_0 \approx 430$ keV cm$^2$.  This 
corresponds to a difference of nearly $1060 \mu K$ at 30 GHz (which is large compared 
to the typical {\it BIMA/OVRO} statistical measurement error of $100-200 \mu K$; Reese et 
al. 2000; 2002).  Of course, there will also be a measurement error associated with 
the cluster mass.  Typically, X-ray-determined masses have associated statistical 
uncertainties of about 20\% (Nevalainen et al. 2000; Finoguenov et al. 2001).  Therefore, 
based on Figure 3, it should be possible to constrain $K_0$ to within about $\pm 100$ keV 
cm$^2$ with current data.

Physically, the diminution in the normalization $y_0-M(r_{500})$ relation can be understood 
since entropy injection greatly reduces the pressure of the cluster gas near the core (and 
therefore $y_0$; see Figure 1) but does not affect its dark matter mass.  It is also 
worth noting that the normalizations of the individual models are a function of redshift.  
For example, the $K_0 = 100$ keV cm$^2$ at $z = 0.2$ predicts a relation somewhat similar 
to that of the $K_0 = 200$ keV cm$^2$ at $z = 0.5$ or the $K_0 = 430$ keV cm$^2$ at $z = 
1.0$.  So, it is extremely important to consider the evolution of galaxy clusters when 
comparisons are made to observational data (i.e., information about the redshift of the 
clusters is required in order to constrain the level of the entropy floor).

Another interesting point is that the difference between the normalizations of 
the models becomes greater at higher redshift.  For example, the difference between 
$y_0$ for the isothermal and $K_0 = 100$ keV cm$^2$ models of a $M(r_{500}) \approx 5.6 
\times 10^{14} M_{\odot}$ cluster at $z = 1.0$ is about 3.7 times larger than at $z = 
0.2$.  This is a result of the fact that the physical size of the isentropic core ($r_c$;
discussed in \S 2) is a larger fraction of the cluster virial radius at higher redshifts.
This trend of increasing $r_c/R_{halo}$ with redshift can be understood as follows.  In 
BBLP02, it was pointed out that there is a critical mass threshold which determines the 
importance of gravitational shock heating.  For clusters with masses above this 
threshold, shock heating significantly increases the entropy of the intracluster gas and 
it begins to dominate the injected non-gravitational entropy (except for in the very 
central regions of the clusters).  For clusters with masses below this threshold, shock 
heating is unimportant.  The critical mass also sets size of the isentropic core, $r_c$.
Incorporating the results of recent high resolution hydrodynamic simulations (Lewis et 
al. 2000), BBLP02 defined the critical mass to be that which gives rise to a gas 
temperature at $R_{halo}$ that is equal to one half of the cluster virial temperature.  
Using this constraint, it is relatively straightforward to show that the critical mass 
(for a fixed entropy floor level) increases with redshift and, therefore, so does the 
ratio $r_c/R_{halo}$.  For gas in the core of the cluster (with entropy equal to $K_0$), 
$P_{core} \propto \rho_{core}^{5/3}$ and, therefore, $T_{core} \propto \rho_{core}^{2/3}$.  
At high redshifts, the cluster gas is denser and, in the case of an $\Omega_m = 1$ 
universe (assumed here for simplicity), scales as $\rho \propto (1+z)^{3}$.  Therefore, 
the temperature of the gas in the core scales as $T_{core} \propto (1+z)^2$.  The gas 
temperature at the virial radius, on the other hand, scales as $T_{halo} \propto (1+z)$.  
Thus, the ratio $T_{core}/T_{halo}$ increases with redshift and in order to match the 
temperature constraint described above, the critical mass threshold and $r_c/R_{halo}$ 
must increase with $z$ as well.   

The steepness of the $y_0-M(r_{500})$ relation is also affected by an entropy floor.  If 
clusters are self-similar, then $y_0 \propto T_X R$, $R \propto T_X^{0.5}$, and 
$M(r_{500}) \propto T_X^{1.50}$ (Evrard et al. 1996).  This leads to 

\begin{equation}
y_0 \propto M(r_{500})
\end{equation}

\noindent while at $z = 0.2$ and 1.0, the $K_0 = 300$ keV cm$^2$ model predicts

\begin{equation}
y_0 \propto M(r_{500})^{1.3,~1.3}
\end{equation}

\noindent A more precise form the $y_0-M(r_{500})$ relation as a function of 
both the entropy floor ($K_0$) and redshift is presented in Table 1 [valid for  
$M(r_{500}) \gtrsim 1.5 \times 10^{14} M_{\odot}$].

The steepening of the relation with the increase in the level of the entropy floor can also 
be understood in an intuitive sense.  It stems from the fact that the relative decrease in 
the gas pressure at the cluster center is strongest for low mass clusters (which have the 
shallowest potential wells; see Figure 1).

Again, because of the increasing importance of entropy injection as one goes to low 
masses, a break from power-law behavior in the $y_0-M(r_{500})$ relation occurs at the 
transition between groups and clusters (not shown).  For entropy floors of $K_0 \gtrsim 
300$ keV cm$^2$, the break becomes quite noticeable for systems with $M(r_{500}) \lesssim 
8 \times 10^{13} M_{\odot}$, which is below the range of system masses that we are most 
interested in and far below the range of system masses that can be observed presently.

We have found that, at least theoretically, the best hope for testing galaxy
clusters for the presence of excess entropy using the $y_0-M(r_{500})$ relation is at 
high redshift, although our results reveal that the relation is sensitive to the entropy
floor level even at low redshifts.  The quantity $M(r_{500})$ is usually determined 
through X-ray observations (e.g., Ettori \& Fabian 1999) and, hence, we have placed it 
in the ``$y_0$-X-ray scaling relations'' subsection.  However, gravitational lensing has 
also been used recently to trace the mass profiles of clusters out to radii near or 
exceeding that of $r_{500}$ with statistical uncertainties similar to that of 
X-ray-determined masses (e.g., Clowe \& Schneider 2001; Gray et al. 2002; Athreya et 
al. 2002).  So, like the  $S_{\nu}-y_0$ relation, the $y_0-M(r_{500})$ relation can 
also potentially be measured completely independent of X-ray results.

\subsubsection{$y_0-T_X$}

In Figure 4, we present scaling relations between $y_0$ and the mean emission-weighted
gas temperature ($T_X$) of a cluster.  The lines hold the same meaning as in Figure 2.
Of the scaling relations presented thus far, the $y_0-T_X$ relation is the most
sensitive to the entropy floor level.  There is a two-fold effect in that entropy
injection diminishes $y_0$ but also {\it increases} $T_X$ for a cluster of mass $M$
(see, e.g., Fig. 1 of MBB02).  For example, at $z = 0.2$, a cluster
that has $T_X \approx 7$ keV and no entropy floor will have a value of $y_0$ that is
about 3.5 times larger than a cluster (also with $T_X \approx 7$ keV) that has an
entropy floor of $K_0 \approx 430$ keV cm$^2$.  At 30 GHz, this corresponds to a
difference of  $\approx 1300 \mu K$.  Current SZ effect and X-ray data can constrain the
central decrement (Compton parameter) and emission-weighted temperature of massive 
clusters
at about the 10\% level (e.g., Reese et al. 2002), making the $y_0-T_X$ relation an
extremely promising test of non-gravitational entropy injection.  Furthermore, the
difference in the normalizations becomes progressively larger with increasing redshift.

The slopes of the predicted relations are also very sensitive to the level of entropy
injection.  Combining $y_0 \propto T_X R$ with $R \propto T_X^{0.5}$, yields the
self-similar result

\begin{equation}
y_0 \propto T_X^{3/2}
\end{equation}

\noindent while at $z = 0.2$ and $1.0$, the $K_0 = 300$ keV cm$^2$ model predicts

\begin{equation}
y_0 \propto T_X^{2.0,~2.1}
\end{equation}

\noindent An analytic expression for the $y_0-L_X$ relation as a function of both the
entropy floor ($K_0$) and redshift is presented in Table 1 (valid for $T_X \gtrsim
3$ keV).

Lastly, we note that a break in slope of the relations occurs at $T_X \lesssim 1$
keV, in good qualitative agreement with 
{\epsscale{1.0}
\plotone{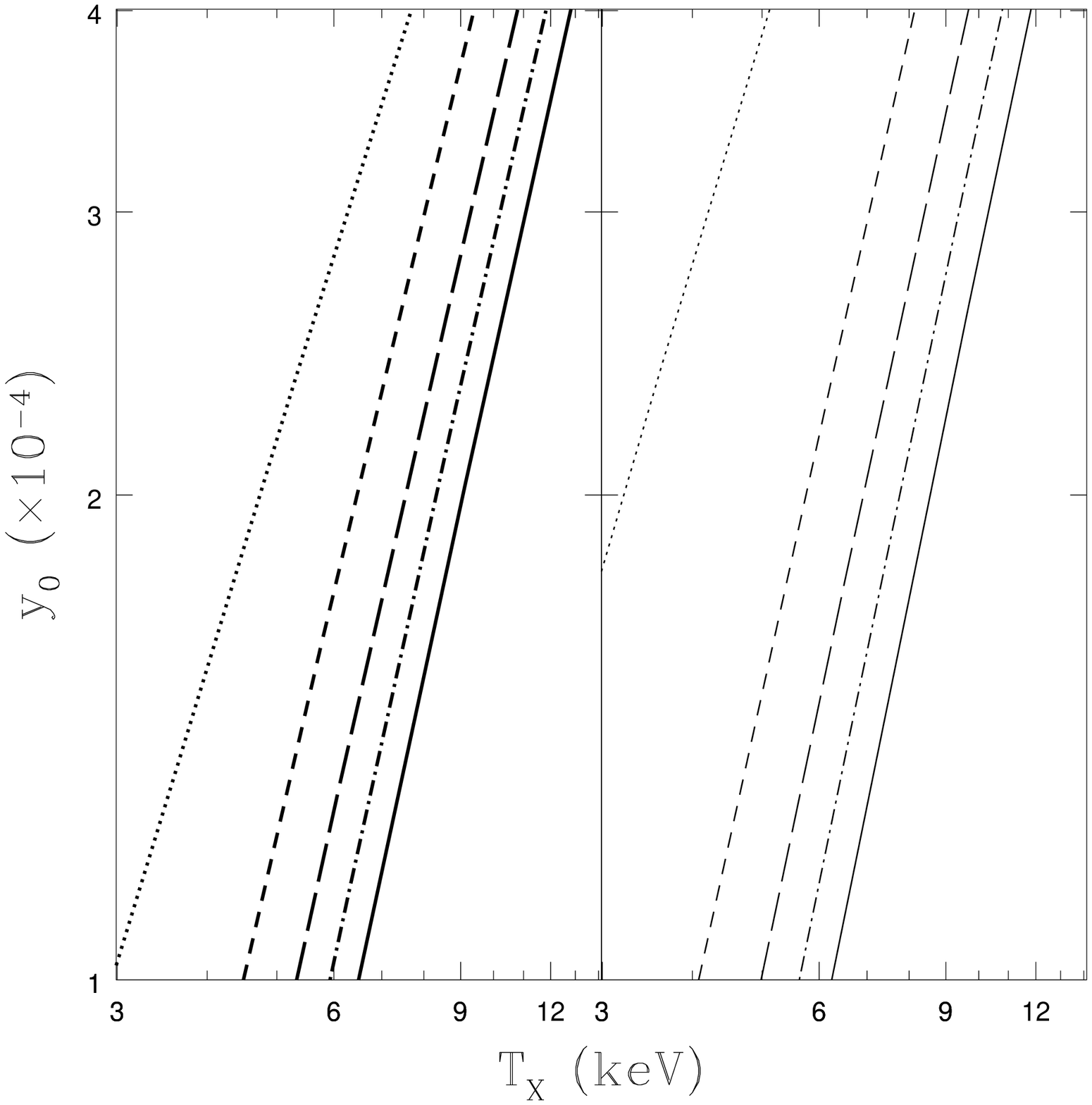}
{Fig. 4. \footnotesize
Comparison of the $y_0 - T_X$ relations.  The lines hold the same meanings as in
Fig. 2.  The left-hand panel is for $z = 0.2$ while the right-hand panel is for $z = 1.0$.
}}
\vskip0.1in
\noindent
the position and steepness of the break found in
the $y-T$ relation of Cavaliere \& Menci (2001; see their Figure 4).  Unfortunately,
present SZ effect observations are essentially limited to systems with $T_X \gtrsim 5$
keV.

\subsubsection{$y_0-L_X$}

In Figure 5, we plot scaling relations between the central $y$ parameter and the 
total bolometric X-ray luminosity ($L_X$) of a cluster.  The lines hold the same meaning 
as in Figure 2.  These relations are more complicated than those presented immediately 
above because both of the coordinates, $y_0$ and $L_X$, are affected by entropy 
injection.  Very interestingly, however, the $y_0-L_X$ relation is almost unchanged by 
the addition of an entropy floor.  Both the self-similar and entropy floor models 
predict very similar slopes and normalizations (at all redshifts).  At first sight, 
this may seem rather surprising.  However, the predicted $y_0-L_X$ relations can be 
understood as follows.  The luminosity of a cluster of fixed mass is reduced by 
entropy injection because $L_X \propto n_e^2$ and entropy injection greatly 
reduces the gas density near the cluster core, where most of the X-ray emission originates.
In the case of the BBLP02 entropy floor models, the result is that $L_X \propto 
M^{1.5}$, roughly (instead of $L_X \propto M^{4/3}$ in the self-similar case).  On the other 
hand, we showed in \S 4.2.1 that an increase in the entropy floor level strongly diminishes 
$y_0$ for a cluster of fixed mass and this leads to $y_0 \propto M^{1.3}$.  This implies 
that entropy injection gives rise to $y_0 \propto L_X^{0.8}$ (roughly), which is very 
close to the self-similar scaling of 

\begin{equation}
y_0 \propto L_X^{3/4}
\end{equation}

\noindent A precise analytic expression for the $y_0-L_X$ relation as a function of both 
the entropy floor ($K_0$) and redshift is presented in Table 1 (valid for $L_X \gtrsim 
10^{44}$ ergs s$^{-1}$).

Despite being insensitive to the entropy floor level, the $y_0-L_X$ relation has at 
least two other potential uses.  
{\epsscale{1.0}
\plotone{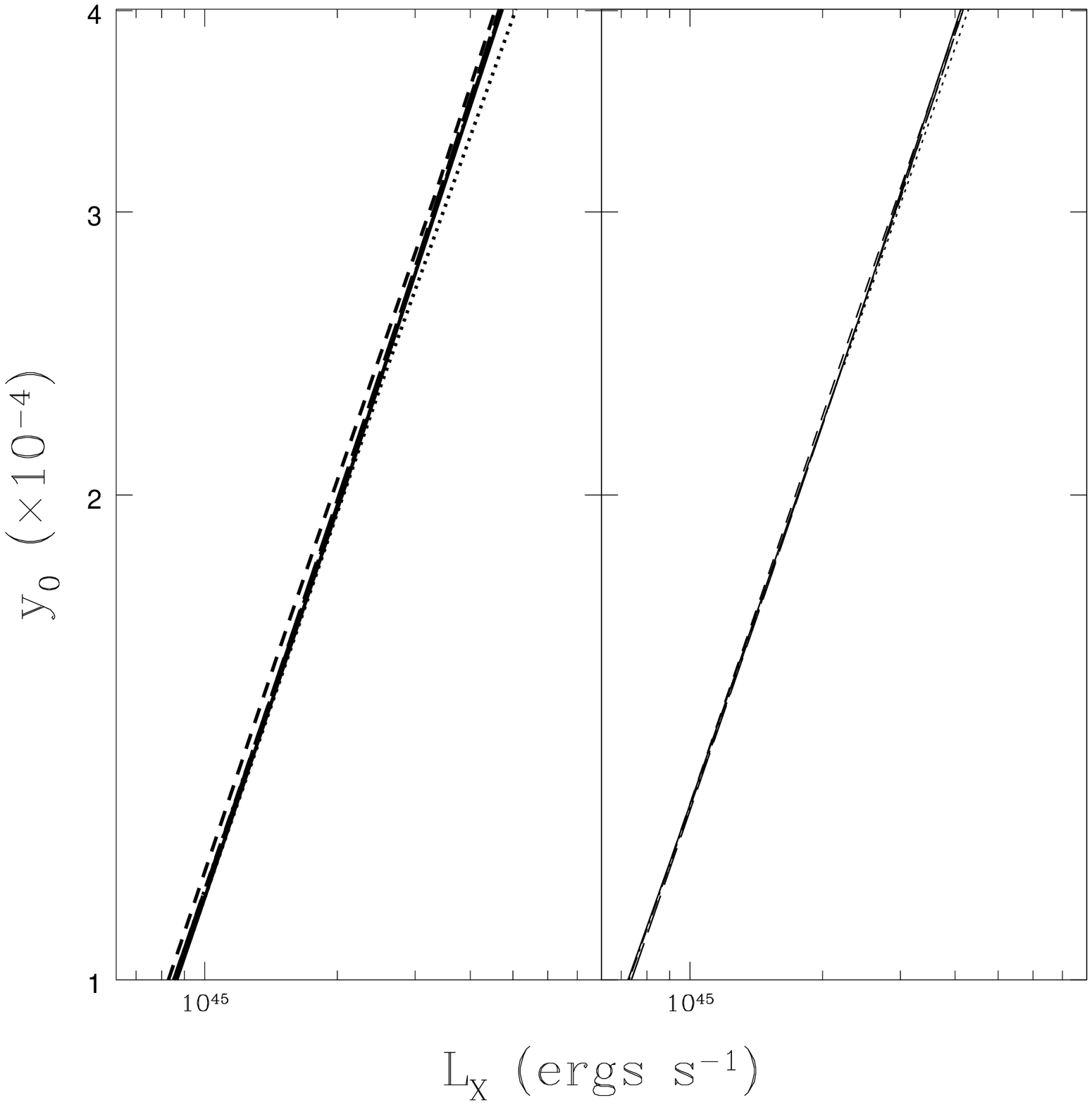}
{Fig. 5. \footnotesize
Comparison of the $y_0 - L_X$ relations.  The lines hold the same meanings as in
Fig. 2.  The left-hand panel is for $z = 0.2$ while the right-hand panel is for $z = 1.0$.
}}
\vskip0.1in
\noindent
First, it could be used as a consistency check of other 
X-ray and SZ effect scaling relations.  Second, it could allow one to deduce the X-ray 
luminosity of clusters in cases where only SZ effect observations are available (or 
vice-versa) without having to worry about the role of ``additional'' gas physics.  The 
accuracy of this will be limited by the relatively large amount of intrinsic scatter present 
in the X-ray luminosities of clusters (see, e.g., studies of the $L_X-T_X$ relation; 
Markevitch 1998; Novicki et al. 2002) and by the measurement errors associated with $y_0$.  

\subsection{$S_{\nu}$-X-ray scaling relations}

Apart for the central $y$ parameter, we also investigated theoretical scaling relations 
between the frequency-independent integrated SZ effect flux density and the X-ray 
observables $M(r_{500})$, $L_X$, and $T_X$.  Again, we evaluated $S_{\nu}/f_{\nu}$ 
within a fixed angular radius of one arcminute and within $R_{halo}$.  Since these 
relations can be reconstructed from Figures 2 - 5 or Table 1, we summarize the 
important points here only.

For the very same reasons discussed in \S 4.2.1, the $S_{\nu}/f_{\nu}-M(r_{500})$ 
relation has its normalization decreased by entropy injection and, also, the relation 
becomes steeper.  How large these effects will be depends upon which radius 
$S_{\nu}/f_{\nu}$ is evaluated within.  Because the dark matter mass is unaffected by 
entropy injection, the $S_{\nu}/f_{\nu}-M(r_{500})$ relation will {\it always} be 
most strongly modified if $S_{\nu}/f_{\nu}$ is evaluated within small projected radii 
(where entropy injection is most prevalent).  Thus, evaluating $S_{\nu}/f_{\nu}$ within 
the central 1 arcminute provides a more sensitive test of the entropy floor than if 
$S_{\nu}/f_{\nu}$ is measured within $R_{halo}$.  Of course, it follows that  
$y_0-M(r_{500})$ relation is the most sensitive to the entropy floor level of any and all of 
the SZ effect-$M(r_{500})$ relations.  The $S_{\nu}/f_{\nu}-M(r_{500})$ 
relation, however, could offer a useful consistency check.

The $S_{\nu}/f_{\nu}-T_X$ relation also becomes steeper and has its normalization
decreased with increasing values of $K_0$ ($S_{\nu}/f_{\nu}$ is diminished while $T_X$
is increased).  The maximum (relative) difference between the various models occurs
when the SZ effect flux densities are measured within small projected radii, although
the increase in $T_X$ is large enough such that the relation is sensitive to the
entropy floor level even when $S_{\nu}/f_{\nu}$ is measured within $R_{halo}$.  Again,
these relations offer useful consistency checks of the more entropy floor-sensitive
$y_0-T_X$ relation.

Finally, entropy injection has an interesting effect on the $S_{\nu}/f_{\nu}-L_X$ 
relation$^7$.  
Both the SZ effect flux density and the luminosity are reduced as the level of entropy 
injected is increased, similar to the $y_0-L_X$ relation.  However, in the case of the 
present relation, the reduction in the luminosity is greater than the reduction in the SZ 
effect flux density (relatively speaking).  This is because $S_{\nu}$ is much less sensitive 
than $y_0$ to the entropy floor level, $K_0$.  The result is that the normalization of the 
$S_{\nu}/f_{\nu}-L_X$ relation is increased by entropy injection.  The sensitivity of this 
relation to the entropy floor level depends on what radius the flux density is evaluated 
within.  It is most sensitive when $S_{\nu}$ is integrated out to $R_{halo}$, since the 
relative reduction in $S_{\nu}$ is at a minimum.  It is least sensitive when the SZ effect 
is evaluated at zero projected radius (i.e., $y_0-L_X$).   

\footnotetext[7]{Note --- there is an error in Figure 12 of the original BBLP02
publication.  The self-similar predictions were incorrectly plotted.}

\section{Discussion \& Conclusions}

Comparisons of observed and predicted cluster X-ray scaling relations has led to 
suggestions that important (non-gravitational) gas physics is being neglected in 
analytic models and numerical simulations.  It has been found that the presence of a core 
in the entropy profiles of clusters (which may be generated through heating 
and/or cooling processes) can account for the deviations between observations and theory 
(e.g., Bower 1997; Balogh et al. 1999; Ponman et al. 1999; Wu et 
al. 2000; Lloyd-Davies et al. 2000; Tozzi \& Norman 2001; Bialek et al. 
2001; Borgani et al. 2001; Voit \& Bryan 2001; BBLP02; MBB02; Voit et al. 
2002; Dav\'{e} et al. 2002; Lloyd-Davies et al. 2002).  So far, X-ray observations have 
provided the only evidence for this excess entropy and it has come from low redshift ($z 
\lesssim 0.2$) clusters alone.  Cosmological dimming is partially responsible for the lack 
of information on the entropy of more distant groups/clusters.  

In the interest of achieving a better physical understanding of important non-gravitational 
processes in high redshift clusters (and how these processes affect cluster formation and 
evolution), we have derived a number of theoretical scaling relations based on SZ effect 
quantities and have analyzed how these relations are modified by the addition of an 
entropy floor.  We find that entropy injection reduces the gas pressure in the cores of 
clusters and, therefore, diminishes the amplitude of both the central and integrated 
Compton parameters (which are directly proportional to the SZ effect).  This translates to 
a steepening of- and a reduction in the normalization of the $y_0-M(r_{500})$ and 
$S_{\nu}/f_{\nu}-M(r_{500})$ relations.  Scaling relations between the Compton parameter(s) 
and X-ray temperature or luminosity are more complicated because both $T_X$ and $L_X$ are 
also affected by entropy injection (whereas the cluster mass $M$ is not).  Because $T_X$ is 
increased by an entropy floor, the $y_0-T_X$ and $S_{\nu}/f_{\nu}-T_X$ relations have their 
normalizations severely reduced by entropy injection.  These relations are also considerably 
steeper than in the case where no entropy floor is present (i.e., the isothermal 
self-similar model).  On the other hand, because $L_X$ is decreased with entropy injection, 
the $y_0-L_X$ relation is almost unaffected by entropy injection, while the normalization 
of the $S_{\nu}/f_{\nu}-L_X$ relation is increased by entropy injection.  Finally, the 
$S_{\nu}/f_{\nu}-y_0$, a relation that can (in principle) be measured entirely through SZ 
effect observations, has its normalization increased by an entropy floor (the relative 
reduction in $y_0$ is much greater than in $S_{\nu}/f_{\nu}$).  The relation is also 
slightly flattened.  Analytic expressions for these relations as a function of entropy floor 
level and redshift can be found in Table 1.

We are aware of only two other studies that have examined the effects of an entropy 
floor on cluster SZ effect scaling relations:  Holder \& Carlstrom (2001) and 
Cavaliere \& Menci (2001).  Using simplified cluster models, Holder \& Carlstrom (2001)
illustrated the potential power of high-$z$ SZ effect observations when it comes 
to studying the physics of ICM.  Their Figure 1 shows the impact of an entropy floor on 
the central SZ effect decrement-cluster core radius relation at $0.5 \leq z \leq 
2.0$.  They find that the addition of an entropy floor tends to decrease the central 
decrement and increase the cluster core radius for clusters of a fixed total mass.  
This is in very good qualitative agreement with our Figure 3, which illustrates that 
the normalization of the $y_0-M(r_{500})$ is decreased by entropy injection, and also 
with our Figure 2, which demonstrates that for clusters of a fixed $y_0$, the SZ effect 
flux density is increased by an entropy floor and, therefore, the SZ effect surface 
brightness becomes less centrally peaked (i.e., the core radius grows).   

Cavaliere \& Menci (2001), on the other hand, implemented a semi-analytic model of 
galaxy formation and clustering and computed a {\it present-day} $y-T_X$ scaling 
relation and analyzed how it was modified by entropy injection from stellar winds and 
supernovae blasts.  In qualitative agreement with the results presented in \S 4, these 
authors found that entropy injection tends to diminish the Compton parameter for 
a cluster of given mean emission-weighted temperature, $T_X$.  However, our theoretical 
results are not directly comparable to those of Cavaliere \& Menci (2001) since their 
focus was mainly on nearby ($z = 0$) groups, whereas we have investigated massive, 
distant clusters such as those regularly observed by the {\it BIMA/OVRO} arrays and the 
{\it Ryle} telescope.  We also point out that the majority of clusters expected to be 
found in upcoming ``blind'' surveys will be at high redshift ($z \sim 1$; Holder et 
al. 2000).  Finally, the Cavaliere \& Menci (2001) cluster models only invoked low levels of 
entropy injection ($K_0 < 100$ keV cm$^2$) consistent with measurements of nearby groups 
(Ponman et al. 1999; Lloyd-Davies et al. 2000), whereas we have generally focused on higher 
levels of entropy injection (which are required to match X-ray observations of nearby 
massive clusters; Tozzi \& Norman 2001; BBLP02; MBB02).

One thing we have not addressed in the present study is the effect of radiative
cooling on SZ effect scaling relations.  Correctly modeling the effects of radiative 
cooling in numerical simulations and analytic cluster models is a daunting task (see for 
Balogh et al. 2001 for a detailed study of the ``cooling crisis'').  Cooling can 
significantly modify not only the ICM density and temperature distributions but it can 
also modify the underlying dark matter distribution of clusters (see Fig. 12 of Lewis et 
al. 2000, for example).  In addition, cooling gives rise to other ``sub-grid'' processes 
such as star formation, outflows, and supernovae explosions.  These, too, will also modify 
the ICM's structure and appearance.  While modeling this is quite difficult, we know from 
both optical and X-ray observations of clusters that cooling must be occurring, at least 
at some level, and, ultimately, must be modeled correctly in order to attain a deep 
physical understanding of the observed properties of clusters.  Numerical simulations and 
analytic models which attempt to take this complicated process into account (e.g., Voit \& 
Bryan 2001; da Silva et al. 2001; Wu \& Xue 2002; Voit et al. 2002; Dav\'{e} et al. 2002; 
White et al. 2002) are apparently able to match many of the features of the X-ray scaling 
relations of nearby clusters.  In this sense, cooling has very similar effects on the ICM 
to those induced by ``preheating''.  Cooling has an advantage over preheating in 
that it does not require some (as of yet) unknown source to transfer large amounts of 
energy into the ICM.  However, cooling alone leads to an expected amount of cooled gas in 
groups and clusters that far exceeds what is observed (Balogh et al. 2001).  Feedback, be 
it through preheating or heating after cluster formation, must also be an important 
process.

We are currently in the process of adding radiative cooling to the BBLP02 analytic models 
(McCarthy et al. in preparation) with the intention of exploring how SZ effect and X-ray 
scaling relations are modified by it.  As far as we know, the effects of cooling on SZ 
effect scaling relations have not yet been examined.  While, in general, we do not expect 
the results of the present study to change significantly with the inclusion of cooling 
(since cooling affects the ICM in a very similar way to that of preheating), in detail, we 
do anticipate subtle (or perhaps not so subtle) differences to be present.  This may also 
lead to a way of separating out the relative contributions of cooling and heating to 
cluster scalings.

Finally, we note that with the recent release of the Reese et al. (2002) data, there are 
now enough SZ effect observations to begin testing the ``entropy floor'' hypothesis by 
mapping out the various correlations involving the SZ and X-ray observables, and 
comparing these to their theoretical counterparts derived in this paper.  It would be 
most interesting to see if these correlations also favor the existence of an entropy 
floor, and whether the level of this floor is comparable to that required to explain 
the X-ray trends.  This will be the focus of McCarthy et al. (2003), a 
companion paper to this one.

\vskip0.25in
We thank the referee for very useful comments and suggestions.  We also thank Peng Oh, 
Kathy Romer, and Mark Voit for helpful discussions.  A. B. would like to acknowledge the 
hospitality extended to him by the Canadian Institute for Theoretical Astrophysics during 
his tenure as CITA Senior Fellow.  I. G. M. is supported by a postgraduate scholarship 
from the Natural Sciences and Engineering Research Council of Canada (NSERC).  A. B. is 
supported by an NSERC operating grant, G. P. H. is supported by the W. M. Keck Foundation, 
and M. L. B. is supported by a PPARC rolling grant for extragalactic astronomy and 
cosmology at the University of Durham.

\newpage

\begin{deluxetable}{ccl}
\tablecaption{Parameters for scaling relations: $Y = A (1+z)^{\gamma} K_2^{\alpha}
X^{\beta}$
\label{tab1}}
\tablewidth{27pc}
\tablehead{
\colhead{Relation: Y - X}                      &
\colhead{}        &
\colhead{parameters}
}
\tablenotetext{~}{$K_2 \equiv K_0/100$ keV cm$^2$.}
\tablenotetext{~}{Note. --- The relations are accurate to better than the $\approx 5\%$ 
level over the ranges $0.1 \lesssim z \lesssim 1$ and $100$ keV cm$^2$ $\lesssim K_0 
\lesssim 700$ keV cm$^2$.}

\startdata
$S_{\nu,arc}/f_{\nu} - y_0$\\
&$A$ & $5.807 \times 10^{4}$\\
&$\gamma$ & $-0.244 -4.812 \log{K_2} + 1.787 (\log{K_2})^2$\\
&$\alpha$ & $0.148 + 0.620 z - 0.007 z^2$\\
&$\beta$ & $1.016 - 0.017 \log{K_2} + 0.150 z - 0.186 z \log{K_2}$\\
\hline
$S_{\nu,halo}/f_{\nu} - y_0$\\
&$A$ & $7.893 \times 10^{7}$\\
&$\gamma$ & $-7.486 - 140.076 \log{K_2} - 0.689 (\log{K_2})^2$\\
&$\alpha$ & $0.527 + 52.220 z - 10.589 z^2$\\
&$\beta$ & $1.478 - 0.269 \log{K_2} - 0.058 z - 0.026 z \log{K_2}$\\
\hline
$y_0 - M(r_{500})$\\
&$A$ & $3.568 \times 10^{-21}$\\
&$\gamma$ & $-1.956 - 88.996 \log{K_2} - 7.675 (\log{K_2})^2$\\
&$\alpha$ & $-3.397 + 35.984 z - 8.293 z^2$ \\
&$\beta$ & $1.130 + 0.243 \log{K_2} + 0.041 z + 0.032 z \log{K_2}$\\
\hline
$y_0 - T_X$\\
&$A$ & $5.668 \times 10^{-6}$\\
&$\gamma$ & $0.188 + 1.287 \log{K_2} - 0.558 (\log{K_2})^2$\\
&$\alpha$ & $-0.734 - 0.451 z + 0.089 z^2$\\
&$\beta$ & $1.880 + 0.287 \log{K_2} + 0.070 z - 0.003 z \log{K_2}$\\
\hline
$y_0 - L_X$\\
&$A$ & $6.659 \times 10^{-41}$\\
&$\gamma$ & $1.504 + 24.922 \log{K_2} + 4.478 (\log{K_2})^2$ \\
&$\alpha$ & $0.887 -11.117 z + 3.057 z^2$\\
&$\beta$ & $0.805 - 0.022 \log{K_2} - 0.005 z - 0.010 z \log{K_2}$\\
\enddata
\end{deluxetable}

\end{document}